\documentclass[final]{ws-ijmpcs}

\begin{document} 

\markboth{V. Gigante \& J. H. Alvarenga Nogueira \& E. Ydrefors \& C. Gutierrez}
{Bethe-Salpeter equation in 2+1 dimensions}

%
\catchline{}{}{}{}{}
%

\title{Study of the homogeneous Bethe-Salpeter equation in the 2+1 Minkowski space\footnote{Bethe-Salpeter equation in 2+1 dimensions}}

\author{V. Gigante}
\address{Laborat\'{o}rio de F\'{i}sica Te\'{o}rica e Computacional - LFTC, Universidade Cruzeiro do Sul, 01506-000, S\~{a}o Paulo, Brazil\\
vitorgigante@yahoo.com.br}

\author{J. H. Alvarenga Nogueira, E. Ydrefors}
\address{Instituto Tecnol\'{o}gico de Aeron\'{a}utica, DCTA, \\S\~{a}o Jos\'e dos Campos, 12228-900, S\~{a}o Paulo, Brazil\\
jorge.ifusp@gmail.com, ydrefors@kth.se}

\author{C. Gutierrez}
\address{Instituto de F\'{i}sica Te\'{o}rica, UNESP, 01156-970, S\~{a}o Paulo, Brazil\\
cristian@ift.unesp.br}

\maketitle

\begin{history}
\received{Day Month Year}
\revised{Day Month Year}
\published{Day Month Year}
\end{history}

\begin{abstract}
The problem of a relativistic bound-state system consisting of two scalar bosons interacting through the exchange of another scalar boson, in 2+1 space-time dimensions, has been studied.  
The Bethe-Salpeter equation (BSE) was solved by adopting the Nakanishi integral representation (NIR) and the Light-Front projection. The NIR allows us to solve the BSE in Minkowski space, which is a big and important challenge, since most of non-perturbative calculations are done in Euclidean space, e.g. Lattice and Schwinger-Dyson calculations. We have in this work adopted an interaction kernel containing the ladder and cross-ladder exchanges. In order to check that the NIR is also a good representation in 2+1, the coupling constants and  Wick-rotated amplitudes have been computed and compared with calculations performed in Euclidean space. Very good agreement between the calculations performed in the Minkowski and Euclidean spaces has been found.  This is an important consistence test that allows Minkowski calculations with the Nakanishi representation in 2+1 dimensions. This relativistic approach will allow us to perform applications in condensed matter problems in a near future.
\keywords{Bethe-Salpeter equation; 2+1 dimensions; Nakanishi representation.}
\end{abstract}

\ccode{PACS numbers: 11.10.St, 03.65.Ge, 03.65.Pm, 11.80.-m}

\section{Introduction}	

Studies of relativistic bound systems in Minkowski space are important in order to understand the structure properties of few-body system in the non-perturbative regime. One of the most important tools for such investigations is the Bethe-Salpeter equation (BSE, [\refcite{Salpeter51}]). Solving the BSE in Minkowski space constitutes a difficult task because of the singularities associated with the propagators and the ones in the interaction kernel. The first successful solution of the Bethe-Salpeter equation in Minkowski space was obtained by Kusaka and Williams [\refcite{Kusaka95}]. They used the Nakanishi Integral Representation (NIR, [\refcite{Nakanishi63}]) and explored the bound-state two-boson problem by using a scalar $\chi^2\phi$ model, i.e.~two massive bosons interacting via the exchange of another massive boson, within the ladder approximation.

 Another important achievement in solving the BSE was obtained by Karmanov and Carbonell  in Ref.~[\refcite{Karmanov06}]. They studied the same model as Kusaka and Williams but used the covariant Light-Front (LF) formalism [\refcite{Carbonell98}] in addition to the NIR. In this way a more convenient solvable framework was acquired and the cross-ladder contribution was subsequently included in the interaction kernel in Ref.~[\refcite{Carbonell06}]. Furthermore, Gigante et al recently used the scalar $\chi^2\phi$ model and studied the cross-ladder effects on the elastic electromagnetic form factor [\refcite{Gigante16}], including the two-body current contribution. The same formalism can also be used to study electroweak form factors in Minkowski space, which is of interest for applications in high-energy neutrino physics, e.g. originated from cosmic rays or remnants~[\refcite{Gaisser05}], which is measured by experiments as ICECUBE~[\refcite{ICECUBE}].
The previously cited applications of the BSE was performed in the ordinary 3+1 dimensional Minkowski space. Recently, in Ref.~[\refcite{Gigante15}], the BSE was also solved for the scalar $\chi^2\phi$  model in 2+1 dimensions within the ladder approximation. In the present work we have extended the calculations of Ref.~[\refcite{Gigante15}] by considering also the cross-ladder contribution to the interaction kernel.   

We have solved the BSE in Minkowski space by using the NIR and the LF projection technique, and in this contribution we show that we can obtain precise coupling constants in comparison with the ones obtained in Euclidean space.  We also obtain the Nakanishi weight functions, which are metric invariant objects and gives the BS amplitudes in Minkowski space, showing that with the Wick-rotated NIR denominator the solution of the BSE matches with the Euclidean BSE solution. The coupling constants and amplitudes have been computed for different values of the mass of the exchanged boson and the two-body binding energy. The computed results are also compared with calculations performed in Euclidean space which were carried out by using the Wick rotation [\refcite{Wick54}].  
This study shows that the NIR is a good representation and can be used to study the relativistic problem in Minkowski space. With this solution, structure observables as parton distributions and form factors, both in timelike and spacelike regions, can be obtained. This is crucial to compute observables that are metric dependent and cannot be computed in Euclidean space. Applications to hadron physics, condensed matter and neutrino physics will be made in a near future.
In 2D materials, for example, there are experiments of light absorption by excitons and trions in monolayers of MoS2 [\refcite{ZhangPRB14}], which has an hexagonal structure and is similar to graphene. 
This kind of materials are known to present Dirac electrons and the excitonic problem could be explored by using the BSE to model the interaction between the electron and the hole.

\section{The Bethe-Salpeter equation and Nakanishi representation}

The BS equation in 2+1 dimensions for a bound state with total momentum $p^2=M^2$ which is composed of two scalars, each of them having mass $m$, reads [\refcite{Gigante15}]
\begin{equation}
\label{Eq:bs_eq}
\Phi(k,p)=\frac{i^2}{\left[(\frac{p}{2}+k)^2-m^2+i\epsilon\right]\left[(\frac{p}{2}-k)^2-m^2+i\epsilon\right]}\int \frac{d^3 k'}{(2\pi)^3}iK(k,k',p;g^2)\Phi(k',p),
\end{equation}
where $K$ denotes the interaction kernel which is given by irreducible Feynman diagrams and $k$ is the relative momentum between the two particles.

The BS amplitude,  can be expressed in terms of the NIR as 
\begin{equation}
\label{Eq:Nakanishi}
\Phi(k,p)=-i\int^{1}_{-1}dz'\int^{\infty}_{0}d\gamma'\frac{g(\gamma',z';\kappa^2)}{\left[\gamma'+m^2-\frac{p^2}{4}-k^2-p\cdot k z'-i\epsilon\right]^{3}},
\end{equation}
where $g(\gamma',z',\kappa^2)$ represents the Nakanishi weight function and the mass of the two-body bound state, $M$, is introduced through the parameter $\kappa^2=m^2-M^2/4>0$. By substituting the expression \eqref{Eq:Nakanishi} for the BS amplitude into  \eqref{Eq:bs_eq} and subsequently integrating over $k^-=k^0-k^3$, one can derive a non-singular generalized integral equation for the weight function. The coupling constant and the weight function can then be determined by using an expansion in terms of orthogonal polynomials and solving the thus obtained generalized eigenvalue problem. For more details we refer to [\refcite{Frederico14}].

\section{Results for the coupling constants and Euclidean amplitudes}

We have in the present work solved the BS equation in 2+1 dimensional Minkowski space by using the method based on an expansion in terms of orthogonal polynomials which was introduced Ref.~[\refcite{Frederico14}]. The solution of the Euclidean BS equation was determined by first performing a Wick rotation of Eq.~\eqref{Eq:bs_eq} and then solving the corresponding integral equation.  

In Table \ref{Tab:coupling_constants} the coupling constants for the ladder plus cross-ladder kernel corresponding to the calculations performed in the Minkowski and Euclidean spaces are compared with each other. In the table we show the results for the representative masses $\mu=0.1\: m$ and $\mu=0.5\: m$ of the exchanged boson, and for various values of the two-body binding energy. It is clearly visible that the Minkowski-space calculations and the Euclidean ones are in perfect agreement with each other.

\begin{table}[!htph]
\tbl{Comparison of the coupling constants computed in the Minkowski and Euclidean spaces. For more information see the text. \label{Tab:coupling_constants}}
{\begin{tabular}{@{}ccccc@{}} \toprule
& \multicolumn{2}{c}{$\mu/m=0.1$} & \multicolumn{2}{c}{$\mu/m=0.5$} \\
 \colrule
 $B/m$ & Mink. & Eucl. & Mink. & Eucl.\\
 \colrule
0.01 & \hphantom{0}0.73 & \hphantom{0}0.72 & \hphantom{0}4.83 & \hphantom{0}4.82 \\
0.1\hphantom{0} & \hphantom{0}3.27 & \hphantom{0}3.27 &  11.98 & 11.98\\
0.2\hphantom{0} & \hphantom{0}5.83 & \hphantom{0}5.83 & 17.26 &  17.26 \\
0.5\hphantom{0} & 13.18 & 13.17 & 29.72 & 29.71 \\
1.0\hphantom{0} & 23.40 & 23.36 & 44.82 & 44.80
\\ \botrule
\end{tabular} \label{ta1}}
\end{table}

\begin{figure}[!htpb]
\centerline{\includegraphics[width=8.0cm]{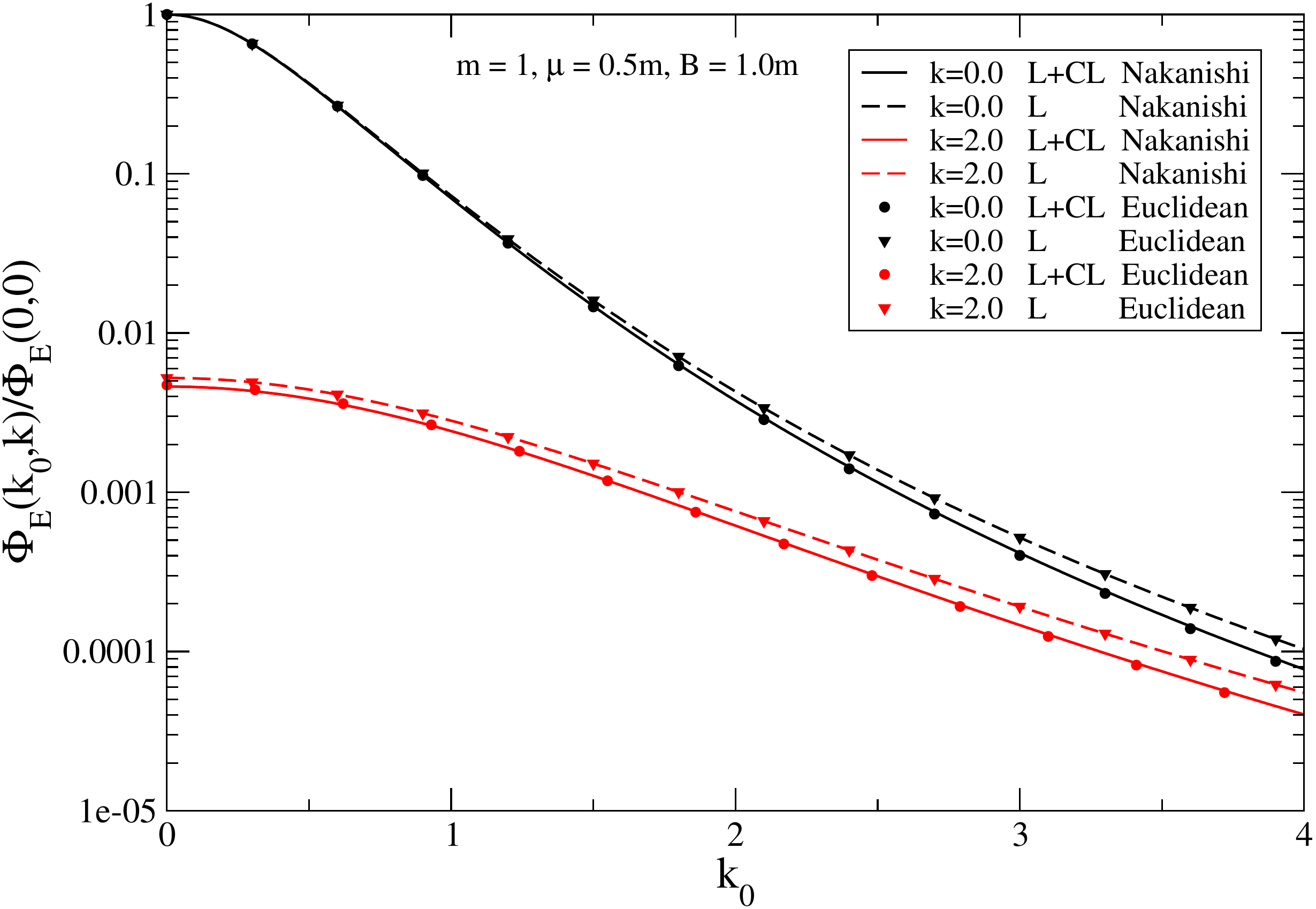}}
\vspace*{8pt}
\caption{Euclidean BS amplitudes as a function of $k^0=k^0_{\text{E}}$ for $k = |\mathbf{k}| = 0.0\:m$ and $k = 2.0\:m$ computed from the Nakanishi weight function (lines) and from the Wick rotated equation
(symbols). Results are shown for ladder (dashed lines) and ladder plus
cross-ladder (solid lines) interaction kernels.  \label{f1}}
\end{figure}

The Euclidean BS amplitude can be computed either by applying a Wick rotation to Eq.~\eqref{Eq:bs_eq} and then solving the ordinary integral equation, or directly from the Nakanishi weight function, see Ref.~[\refcite{Frederico16}] for more details. In Fig.~\ref{f1}  the Euclidean BS amplitude as a function of $k^0_{\text{E}}$ computed by the two mentioned methods is shown. Results are displayed for the ladder and the ladder plus cross-ladder kernels. It is seen in the figure that the results obtained by using the two methods are indistinguishable from each other.

\section*{Acknowledgments}

We thank for the financial support from Conselho Nacional de Desenvolvimento Cient\'ifico e Tecnol\'ogico(CNPq), Coordena\c c\~ao de Aperfei\c coamento de Pessoal de N\'ivel Superior 
(CAPES) and Funda\c c\~ao de Amparo \`a Pesquisa do Estado de S\~ao Paulo (FAPESP). J. H. A. N. acknowledges 
the support of the grant $\#$2014/19094-8 from S\~ao Paulo Research Foundation (FAPESP).


\begin{thebibliography}{0}    
\bibitem{Salpeter51} E. E. Salpeter and H. A. Bethe, {\it Phys. Rev.} {\bf 84}, 1232 (1951).
\bibitem{Kusaka95} K. Kusaka and A. G. Williams, {\it Phys. Rev. D} {\bf 51}, 7026 (1995); K. Kusaka, K. Simpson and A. G. Williams, {Phys. Rev. D} {\bf 56} 5071 (1997).
\bibitem{Nakanishi63}  N. Nakanishi, {\it Graph Theory and Feynman Integral} (Gordon and Breach, New York, 1971).
\bibitem{Karmanov06} V. A. Karmanov and J. Carbonell, {\it Eur. Phys. J. A} {\bf 27}, 1 (2006). 
\bibitem{Carbonell98} J. Carbonell et al, {\it Phys. Rep.} {\bf 300}, 215 (1998).
\bibitem{Carbonell06} J. Carbonell and V. A. Karmanov, {\it Eur. Phys. J. A} {\bf 27}, 11 (2006).
\bibitem{Gigante16} V. Gigante et al, Phys. Rev. D 95, 056012 (2017). 
\bibitem{Gaisser05} T.~K. Gaisser, Physica Scripta {\bf T121}, 51 (2005).
\bibitem{ICECUBE} {ICECUBE-South Pole Neutrino Observatory}, {https://icecube.wisc.edu/}.
\bibitem{Gigante15} V. Gigante et al, {\it Few-Body Syst.} {\bf 56}, 375 (2015). 
\bibitem{Wick54} G. C. Wick, {\it Phys. Rev.} {\bf 96}, 1124 (1954).
\bibitem{ZhangPRB14} C. Zhang, H. Wang, W. Chan, C. Manolatou, and F. Rana, Phys. Rev. B {\bf 89}, 205436 (2014).
\bibitem{Frederico14} T. Frederico, G. Salm{\`{e}} and M. Viviani, {\it Phys. Rev. D} {\bf 89}, 016010 (2014).
\bibitem{Frederico16} T. Frederico et al, {\it Few-Body Syst.} {\bf 57}, 549 (2016).  
\end{thebibliography}
\end{document}